\documentclass[10pt,technote,compsoc]{IEEEtran}

\ifCLASSOPTIONcompsoc
  \usepackage[nocompress]{cite}
\else
  \usepackage{cite}
\fi

\ifCLASSINFOpdf
\else
\fi

\usepackage[utf8]{inputenc} % allow utf-8 input
\usepackage[T1]{fontenc}    % use 8-bit T1 fonts
\usepackage[hidelinks]{hyperref}
\usepackage{url}            % simple URL typesetting
\usepackage{booktabs}       % professional-quality tables
\usepackage{amsfonts}       % blackboard math symbols
\usepackage{nicefrac}       % compact symbols for 1/2, etc.
\usepackage{microtype}      % microtypography
\usepackage{xcolor}         % colors

\usepackage{graphicx}  
\usepackage{epstopdf}
\usepackage{hyperref}
\usepackage{multirow}
\usepackage{multicol} 
\usepackage{arydshln}
\usepackage{booktabs}
\usepackage{amsmath}

\usepackage{wrapfig}

\begin{document}
\title{GradMDM: Adversarial Attack on Dynamic Networks}

% The \author macro works with any number of authors. There are two commands
% used to separate the names and addresses of multiple authors: \And and \AND.
%
% Using \And between authors leaves it to LaTeX to determine where to break the
% lines. Using \AND forces a line break at that point. So, if LaTeX puts 3 of 4
% authors names on the first line, and the last on the second line, try using
% \AND instead of \And before the third author name.

% \author{%
%   David S.~Hippocampus\thanks{Use footnote for providing further information
%     about author (webpage, alternative address)---\emph{not} for acknowledging
%     funding agencies.} \\
%   Department of Computer Science\\
%   Cranberry-Lemon University\\
%   Pittsburgh, PA 15213 \\
%   \texttt{hippo@cs.cranberry-lemon.edu} \\
%   % examples of more authors
%   % \And
%   % Coauthor \\
%   % Affiliation \\
%   % Address \\
%   % \texttt{email} \\
%   % \AND
%   % Coauthor \\
%   % Affiliation \\
%   % Address \\
%   % \texttt{email} \\
%   % \And
%   % Coauthor \\
%   % Affiliation \\
%   % Address \\
%   % \texttt{email} \\
%   % \And
%   % Coauthor \\
%   % Affiliation \\
%   % Address \\
%   % \texttt{email} \\
% }

%together
\author{Jianhong Pan, Lin Geng Foo, Qichen Zheng, Zhipeng Fan, Hossein Rahmani, Qiuhong Ke, Jun Liu
\IEEEcompsocitemizethanks{\IEEEcompsocthanksitem J. Pan, L. G. Foo, Q. Zheng and J. Liu are with SUTD. Z. Fan is with NYU. H. Rahmani is with Lancaster University. Q. Ke is with Monash University.
(J. Pan and L. G. Foo contributed equally.)%\protect\\
% E-mail: see http://www.michaelshell.org/contact.html
%\IEEEcompsocthanksitem Jianhong Pan and Lin Geng Foo contributed equally.
}% <-this % stops an unwanted space
}

% \markboth{Journal of \LaTeX\ Class Files,~Vol.~14, No.~8, August~2015}%
% {Shell \MakeLowercase{\textit{et al.}}: Bare Demo of IEEEtran.cls for Computer Society Journals}

\markboth{TPAMI - Short Paper Submission}%
{Shell \MakeLowercase{\textit{et al.}}: Bare Demo of IEEEtran.cls for Computer Society Journals}

\IEEEtitleabstractindextext{%
\begin{abstract}
 Dynamic neural networks can greatly reduce computation redundancy without compromising accuracy by adapting their structures based on the input. 
 In this paper, we explore the robustness of dynamic neural networks against \textit{energy-oriented attacks} targeted at reducing their efficiency.
 Specifically, we attack dynamic models with our novel algorithm GradMDM.
 GradMDM is a technique that adjusts the direction and the magnitude of the gradients to effectively find a small perturbation for each input, that will activate more computational units of dynamic models during inference. 
 We evaluate GradMDM on multiple datasets and dynamic models, where it outperforms previous energy-oriented attack techniques, significantly increasing computation complexity while reducing the perceptibility of the perturbations. 
\end{abstract}

% \begin{IEEEkeywords}
% Computer Society, IEEE, IEEEtran, journal, \LaTeX, paper, template.
% \end{IEEEkeywords}
}

\maketitle

\IEEEdisplaynontitleabstractindextext
\IEEEpeerreviewmaketitle

\renewcommand{\baselinestretch}{0.9}

\section{Introduction}

\IEEEPARstart{I}n recent years, research on Deep Neural Networks (DNNs) has seen tremendous progress, with DNNs being widely developed for various areas, such as image classification \cite{he2016deep, huang2017densely, dosovitskiy2020image, liu2021swin, liu2022convnet},
segmentation \cite{long2015fully, ronneberger2015u, badrinarayanan2017segnet, chen2017deeplab, he2017mask} 
and object detection \cite{girshick2015fast, liu2016ssd, lin2017focal, he2017mask, carion2020end}. 
However, as DNNs continue to attain better and better performance, they have also been steadily increasing in size, even reaching up to the scale of billions of parameters \cite{vaswani2017attention}. 
These computation-heavy models are highly challenging to be deployed on embedded and mobile devices, motivating research on the development of efficient models \cite{cheng2017survey, cheng2018recent} to accelerate the inference process.

To this end, dynamic neural networks \cite{han2021dynamic} offer a promising strategy for accelerating the inference process, 
by skipping the redundant computations on-the-fly. This is typically achieved by introducing gating mechanisms \cite{han2021dynamic} to adaptively execute only a subset of computational units, conditioned on the input sample. Thus, dynamic neural networks provide a better trade-off between accuracy and efficiency than their static counterparts \cite{han2021dynamic}, as shown throughout many works \cite{Wang_2018_ECCV,figurnov2017spatially,tang2021manifold}.

Despite the appeal and rising popularity of dynamic neural networks, the robustness of their mechanisms and efficiency gains is still an important issue that requires more studies.
As shown in ILFO \cite{Haque_2020_CVPR}, dynamic neural networks are highly susceptible to \textit{energy-oriented attacks}, which are adversarial attacks aiming at boosting the energy consumption and computational load of the dynamic neural network.
These energy-oriented attacks can drastically reduce the number of FLOPs (floating-point operations) saved by the dynamic mechanisms.
Importantly, such research on energy-oriented attacks broadens our understanding of dynamic mechanisms,
and contributes an alternative point-of-view studying the potential risks of dynamic networks.

In this paper, we provide insights on energy-oriented attacks and identify two key pitfalls of the existing method \cite{Haque_2020_CVPR}, as shown in Sec. \ref{sec:RectifiedMag} and Sec. \ref{sec:RectifiedDirection}. 
Firstly, ILFO \cite{Haque_2020_CVPR} attacks gates to lower a loss function (a Complexity Loss),
but the direct minimization of this loss might not be enough to fully maximize the number of activated gates for a given ``allowable" amount of perturbation to the image.
Secondly, we observe difficulties in the joint optimization of the many gates' losses, that makes it difficult to simultaneously keep many gates activated.
We attribute this to the presence of \textit{conflicting gradients} between the gates, which makes it hard to activate new gates without interfering with already-activated gates.

To mitigate the above-mentioned issues, we propose GradMDM, a \textbf{Grad}ient \textbf{M}agnitude and \textbf{D}irection \textbf{M}anipulation method for energy-oriented attacks against dynamic networks. 
(1) We introduce a Power Loss to modify the magnitudes of the losses across various gates.
This loss prioritizes the optimization of gates in a principled manner -- to find an optimal spot with more activated gates along the Pareto Frontier -- which allows for activation of more gates while keeping the input perturbation imperceptible.
(2) Next, we propose Complexity Gradient Masking (CGM), a technique that adjusts the directions of the Complexity Gradient 
to reduce gradient conflicts among gates.
It involves a gradient projection process to rectify the gradient directions between activated and deactivated gates such that there are less conflicts between them, thus making it easier to activate new gates without negatively affecting already-activated gates. 
(3) We demonstrate the efficacy of GradMDM on multiple dynamic neural networks across various datasets, where GradMDM effectively increases computations with less perceptible perturbations across all settings.

\section{Related Work}

\textbf{Dynamic Neural Networks} 
\cite{veit2018convolutional,Wang_2018_ECCV,yang2019condconv,li2020learning,huang2018multiscale,graves2016adaptive,figurnov2017spatially,tang2021manifold,liu2018dynamic,wu2018blockdrop,li20212d,wang2021not,yin2022avit,kaya2019shallow} 
achieve efficiency while maintaining good accuracy by dynamically adjusting model architectures to allocate appropriate computation conditioned on each input sample.  
This reduces redundant computations on those ``easy'' samples, improving the inference efficiency \cite{han2021dynamic}.
\textit{Dynamic depth networks} \cite{graves2016adaptive,figurnov2017spatially,kaya2019shallow,Wang_2018_ECCV} adjust their architectures to use only activated layers to conduct inference, and often achieve efficient inference in one of two ways: early termination or conditional skipping.
\textit{Early termination networks} \cite{graves2016adaptive,figurnov2017spatially,kaya2019shallow} are able to terminate their inference process early.
For instance, Adaptive Computation Time (ACT) \cite{graves2016adaptive} augments an RNN with a ``halting score"  
that adaptively terminates the recurrent computations to reduce cost.
SACT \cite{figurnov2017spatially} extends this concept to ResNets for vision tasks by applying ACT to each residual block group, allowing for an early exit.
Another early termination network is the Shallow-deep network \cite{kaya2019shallow} which uses internal classifiers to facilitate confidence-based early exits.
On the other hand, \textit{conditional skipping models} \cite{Wang_2018_ECCV,wu2018blockdrop,veit2018convolutional} can decide to skip layers or blocks based on each input.
SkipNet \cite{Wang_2018_ECCV} assigns a gating module to each convolutional block in CNNs, which decides whether to execute or skip it, and is trained via reinforcement learning.
\textit{Dynamic width networks} \cite{tang2021manifold,yin2022avit,hua2019channel} selectively activate multiple components within the same layer, such as channels and neurons, based on each instance. 
Early studies \cite{bengio2013estimating,cho2014exponentially,bengio2015conditional} achieve dynamic width by adaptively controlling the activation of neurons or parameters, e.g., via stochastic gating units \cite{bengio2013estimating,cho2014exponentially}.
Subsequently, some works explore dynamic channel pruning in CNNs
\cite{tang2021manifold,lin2017runtime,hua2019channel}, 
which selectively activates convolutional channels.
For instance, ManiDP \cite{tang2021manifold} uses the relationship between samples (i.e., manifold information) to identify and dynamically remove redundant channels or filters.

In this paper, we propose a method to generate adversarial samples to perform energy-oriented attacks. 
We showcase the efficacy of our method on both the dynamic depth structures (SkipNet \cite{Wang_2018_ECCV} and SACT \cite{figurnov2017spatially}) as well as the dynamic width structures (ManiDP \cite{tang2021manifold}).

\textbf{Energy-oriented Attacks} \cite{Haque_2020_CVPR,Hong2021DeepSloth,pan2022gradauto} focus on delaying the inference speed of the target dynamic model, by incurring more computations within the model and increasing the energy consumption of inference.
A successful energy-oriented attack will leave the dynamic neural network no longer efficient,
therefore defeating the purpose of using them in the first place.
\nocite{pan2022gradauto}
Despite its significance, research into adversarial attacks on dynamic neural networks is quite limited. 
ILFO \cite{Haque_2020_CVPR} is the first work investigating the robustness of dynamic neural networks. 
DeepSloth \cite{Hong2021DeepSloth} attacks the early termination-based dynamic neural nets with an adversarial example crafting technique.

In this work, we show that directly searching for adversarial examples based on gradients, as done in previous works, could be sub-optimal.
To address this, we propose two remedies to rectify the gradients, leading to improved adversarial performance with less perceptible perturbations.

\section{Background}

\textbf{Gate Mechanism.}
In many works \cite{Wang_2018_ECCV,veit2018convolutional,leroux2018iamnn,guo2019dynamic,yu2019any,jin2020adabits,shen2020fractional,li2021dynamic} with dynamic depth or dynamic width architectures, a gating network $G_i(\cdot)$ with a sigmoid activation function is often used to generate gating values for gate $g_i$ as follows: 
\begin{align}
    g_i=\left \{
        \begin{array}{ll}
          1,     & G_i(\boldsymbol{x}) \geq \tau, \\
          0 ,     & G_i(\boldsymbol{x})<\tau,
        \end{array}
      \right.
\end{align}
where $G_i(\boldsymbol{x})\in [0,1]$ is the estimated gating value of the $i^{th}$ gate based on the input $x$, and $\tau\in (0,1)$ denotes the threshold. 
If $g_i =1$, the gate will execute the optional computations, and if $g_i =0$, the gate will skip the optional computations.
We remark that $g_i$ can generally refer to the gating value of a gate in a dynamic depth network (e.g., for layer activation) or a dynamic width network (e.g., for channel activation), and so attacks on $g_i$ can work for both dynamic depth and dynamic width architectures.

When performing an energy-oriented attack, our task is generally to make as many $g_i$ produce values of 1 as possible.
This is done by perturbing the input sample $x$ to raise the gating values to go beyond the threshold $\tau$, which leads to the corresponding gates being \textit{activated}, incurring additional computational cost to the dynamic neural networks. 
To clarify, in this work, we take ``activated'' gates (i.e., $g_i = 1$) to mean that the gate is executing the optional computations and incurring higher computational costs.

\section{Method: GradMDM Attack}

In this section, we first formulate a Baseline Loss (Sec.~\ref{The Overall Objective for Computational Complexity Attack}) from the ground up, which
can be used to conduct energy-oriented attacks against dynamic architectures.
Then, we propose two techniques involved in our GradMDM attack that can improve upon the Baseline Loss.
\textbf{1)} For the sake of achieving a better result, the attacks on some gates might be more important than others.
Thus, we introduce a Power Loss (Sec.~\ref{sec:RectifiedMag}) that prioritizes the attacking of some gates over others to maximize the number of activated gates.
\textbf{2)} As it might be difficult to jointly optimize many gates simultaneously (i.e., gradient conflicts might occur between different gates' losses), we propose our Complexity Gradient Masking (CGM) method (Sec.~\ref{sec:RectifiedDirection}) to rectify the directions of gradients.

\subsection{Ground Up Formulation of Baseline Loss for Computational Complexity Attack}
\label{The Overall Objective for Computational Complexity Attack}

Following previous works \cite{Haque_2020_CVPR,Hong2021DeepSloth}, we construct an adversarial sample by creating a human-imperceptible perturbation to modify a given input.
By conducting iterative updates using a carefully designed objective function, perturbations can be optimized to alter the predictions (i.e. gating values) of the threatened dynamic models and invalidate their acceleration strategy.
We detail this kind of attack below.

\noindent \textbf{Computational Complexity Attack.}
First, we initialize a specific perturbation $\boldsymbol{\delta}\in \mathbb{R}^{3\times H\times W}$ and use it to modify the input image as: $\boldsymbol{x}^{\prime}_{0}=\boldsymbol{x}_{0}+\boldsymbol{\delta}$,
where $H,W$ denote the height and width of the input image.
$\boldsymbol{x}_{0},\boldsymbol{x}^{\prime}_{0}\in [0,1]^{3\times H\times W}$ respectively denote the original input and modified input. 

We then follow \cite{carlini2017towards} 
to use $\text{tanh}(\cdot)\in [-1,1]$, so that the pixel values of the modified input $\boldsymbol{x}^{\prime}_{0}$ are within $[0,1]$ similar to the original input $\boldsymbol{x}_{0}$, as follows:
\begin{align}
    \label{eq:tanh_mapping}
    \boldsymbol{x}^{\prime}_{0}=
    \frac{1}{2}\cdot
    (\text{tanh}(\boldsymbol{x}_{0}+\boldsymbol{\delta})+1).
\end{align}

To optimize the perturbation $\boldsymbol{\delta}$ for increasing the complexity of the network, we refer to Szegedy \textit{et al.}~\cite{szegedy2013intriguing} and formally define the objective and constraints as:
\begin{align}
    \label{eq.OriginalObjective}
    &\operatorname{minimize}_{\boldsymbol{\delta}}\left\lVert \frac{1}{2}\cdot
    (\text{tanh}(\boldsymbol{x}_{0}+\boldsymbol{\delta})+1)-\boldsymbol{x}_{0} \right\rVert_2^2 \\
    &~~s.t.~~ G_{i}(\boldsymbol{x}_0, \boldsymbol{\delta}) \ge \tau,
\end{align}
where $\|\cdot\|_2$ denotes the $L2$-norm, and its objective is to minimize the deviation between the original input and the modified input to prevent them from being easily distinguishable. 
The constraint enforces the gating value $G_i$ to be above the threshold, to activate the execution of the $i^{\text{th}}$ gate for more computational complexity. 
Here, we use $G_i(\boldsymbol{x}_{0}, \boldsymbol{\delta})$ as the gating function to indicate that it depends on both the input and the perturbation.

\noindent \textbf{Baseline Loss.}
As it is difficult to optimize under the constrained formulation in Eq. \ref{eq.OriginalObjective}, we can instead optimize our perturbations $\boldsymbol{\delta}$ using a loss $\mathcal{L}_{\text{relaxed}}$ derived from the relaxation of Eq. \ref{eq.OriginalObjective}:
\begin{equation}
\begin{aligned}
    &\mathcal{L}_{\text{relaxed}} (\boldsymbol{x}_{0},\boldsymbol{\delta}) 
    \\&= \gamma \left\lVert \frac{1}{2}\cdot
    (
        \text{tanh}(\boldsymbol{x}_{0}+\boldsymbol{\delta})+1
    )-\boldsymbol{x}_{0}  \right\rVert_2^2
    -\sum_{i=1}^N  ( G_i(\boldsymbol{x}_0,\boldsymbol{\delta}) - \tau) ,
    \label{Eq.LagrangianRelaxation}
\end{aligned}
\end{equation}
where $\gamma$ is the weight of the first term, and $N$ represents the total number of gates in the dynamic network.
$\mathcal{L}_{\text{relaxed}}$ in Eq. \ref{Eq.LagrangianRelaxation} can be applied as a loss on the gate outputs, and backpropagated to optimize the perturbations $\boldsymbol{\delta}$.
Moreover, the second term in Eq. \ref{Eq.LagrangianRelaxation} penalizes the outputs of \textit{all} gates $\{G_i\}_{i=1}^N$.
To further streamline the loss, we modify the second term such that losses only apply to gates $G_i(\boldsymbol{x}_0,\boldsymbol{\delta})<\tau$. Thus, the Baseline Loss $\mathcal{L}_{\text{base}}$ is as follows:
\begin{equation}
\begin{aligned}
    &\mathcal{L}_{\text{base}} (\boldsymbol{x}_{0},\boldsymbol{\delta}) = \gamma \mathcal{L}_{\text{MSE}} + \mathcal{L}_{\text{C}}
    \\
    &=
    \gamma \left\lVert \frac{1}{2}\cdot
    (
        \text{tanh}(\boldsymbol{x}_{0}+\boldsymbol{\delta})+1
    )-\boldsymbol{x}_{0} \right\rVert_2^2
    -\sum_{i=1}^N \min(G_i(\boldsymbol{x}_0,\boldsymbol{\delta}),\tau), 
    \label{Eq.StationarityFinal}
\end{aligned}
\end{equation}
which optimizes only the deactivated gates, so that the loss can ignore already-activated gates, and thus focuses more on minimizing the image perturbation. 
The first term $\mathcal{L}_{\text{MSE}}$ can be considered to be an Imperceptibility Loss which minimizes the visibility of perturbations, while the second term $\mathcal{L}_{\text{C}}$ is a Complexity Loss that maximizes the computational complexity of the dynamic network by activating the gates.
We note that ILFO \cite{Haque_2020_CVPR} uses the same mechanism as Eq.~\ref{Eq.StationarityFinal} to perform energy-oriented attacks.

\subsection{Rectified Loss Magnitudes}
\label{sec:RectifiedMag}

An issue with the Baseline Loss is that directly optimizing using the Complexity Loss might not result in the maximum number of gates activated,
because the Complexity Loss fails to consider the \textit{overall optimal balance between the gating values} during optimization of this loss. 
Below, we first explain the problem in more detail by treating our task as a multi-objective optimization problem. Then, through the lens of the Pareto Frontier \cite{jahan2016multi}, 
we propose a Power Loss that can balance the optimization of gating values along the boundary such that more gates can be activated.

\begin{figure*}[tp]
    \center
    \includegraphics[width=\textwidth]{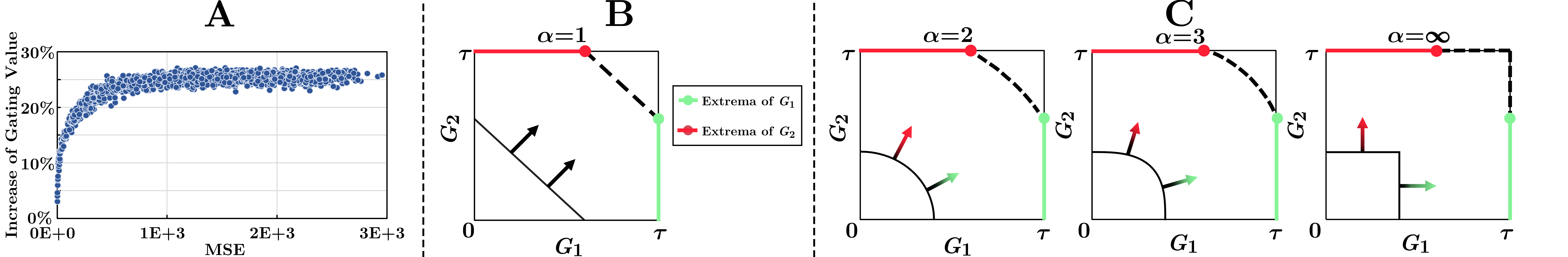}      
    \caption{
    A: Plot showing the trade-off between the two optimization objectives.
    MSE denotes the value of $L_\text{MSE}$.
    B: Visualization of loss contours of two gating values, where gradients are represented by arrows.
    C: Visualization of loss contours with different $\alpha$ for $\mathcal{L}_{\text{P}}$.
    Higher $\alpha$ leads to larger deformation of the contour lines, with the gradient arrows pointing away
    from the dotted line.
    }
    \label{Figure.VisualizationOfGradient}
    \vspace{-0.5cm}
\end{figure*}

\noindent \textbf{Pareto Frontier.}
In multi-objective optimization, a Pareto Frontier refers to a set of optimal solutions, where no single objective can be optimized further without making another objective worse off.
In our scenario, as we are simultaneously optimizing for the minimization of perturbations (Imperceptibility Loss), as well as maximizing the attack efficacy (Complexity Loss), 
the Pareto Frontier represents the optimal boundary along which we balance these objectives.
In practice, we can roughly assume that, when trained to convergence, the input perturbations will achieve a pair of objective values that lies close to this Pareto Frontier, which we verify experimentally through rigorous training and testing using the Baseline Loss defined in Eq.~\ref{Eq.StationarityFinal}.
Results are shown in Fig \ref{Figure.VisualizationOfGradient}.A, where we plot values of the two optimization objectives at convergence: 
the Complexity objective versus the Imperceptibility objective.
Each point represents the result of an evaluation of the attack on the ImageNet dataset, and
different points are obtained by varying the value of $\gamma$ 
that controls the trade-off between the two objectives.
We obtain pairs of objective values that quite neatly form a boundary -- which we argue to be approximately the Pareto Frontier of our two objectives.

This observed Pareto Frontier implies that, for a \textit{given size of perturbations}, 
we can only expect the overall Complexity Loss to be optimized to a certain limited extent (on the Pareto Frontier) after training.
If we want to further optimize the Complexity Loss past that point, it can be done by tuning $\gamma$, but doing so may come at the expense of a higher MSE (i.e., the  Imperceptibility Loss).
However, we would like to activate more gates, without sacrificing our MSE objective.
To do this, we seek to find an optimal balance among the gating values, that pushes more gating values over the threshold $\tau$.

To explore this idea in more detail, we elaborate using a simple example with 2 deactivated gates, and
visualize the loss contours based on their gating values $G_1$ and $G_2$ in Fig. \ref{Figure.VisualizationOfGradient}.B. 
In order to activate these gates, gating values $G_1$ and $G_2$ need to reach $\tau$ or higher, which represents the area beyond the right and top edge of the figure, respectively.
When optimizing these gates using the Complexity Loss $\mathcal{L}_{\text{C}}$, we get a loss of $-\min(G_1,\tau) - \min(G_2,\tau) = -(G_1 + G_2)$.
Since $-(G_1 + G_2)$ is linear, the loss contours are straight lines.
In order to minimize the loss at a point $(G_1,G_2)$, we can go in the direction of the gradient $-\nabla [-(G_1 + G_2)] = (\frac{\partial (G_1 + G_2) }{\partial G_1} , \frac{\partial (G_1 + G_2) }{\partial G_2} )^T = (1,1)^T$,
where $(1,1)^T$ is a vector indicating the direction on the loss contour map, visualized as arrows pointing to the top right at 45$^\circ$.

In order to keep a good image quality and maintain a low MSE, there is a certain limit to how much we can optimize the loss $-(G_1 + G_2)$, which sets a \textit{soft limit} on how high the sum of gating values can get.
This can be represented by a 
``budget'' $B$, visualized as a dotted line $-(G_1 + G_2) = B$ in Fig. \ref{Figure.VisualizationOfGradient}.B.
This implies that top right corner beyond the dotted line is inaccessible if we want to keep the image quality at a certain good level.
Moreover, if we use the Complexity Loss, many gradient arrows will be directed towards the dotted line, where none of the gating values can reach $\tau$.
To improve upon this, we aim to change the shape of loss contours
such that at least one of the gating values will reach $\tau$.
This allows us to keep perturbations to a minimum (i.e., staying within budget $B$), while driving the optimization to maximize the number of activated gates.

\noindent \textbf{Power Loss.}
Based on our insights, we propose a Power Loss which can improve the shape of the loss landscape, by \textit{pulling the contour lines outwards},
allowing us to optimize towards a spot that activates more gates.
We rewrite the Complexity Loss $\mathcal{L}_\text{C}$ into a Power Loss $\mathcal{L}_\text{P}$ as:
\begin{align}
    \mathcal{L}_\text{P}(\boldsymbol{x}_{0},\boldsymbol{\delta}) &=- \sum_{i=1}^N  \min(G_i(\boldsymbol{x}_0,\boldsymbol{\delta}),\tau)^\alpha,     
    \label{eq.power_loss}
\end{align}
where $\alpha\in [1,+\infty)$ can be used to control the deformation of the contour line. Note that, for the case of $\alpha=1$, the loss becomes the original Complexity Loss.
For deactivated gates, the loss contour lines can be formulated as $-\min(G_1,\tau)^\alpha - \min(G_2,\tau)^\alpha = -(G_1^\alpha + G_2^\alpha) = d$. As $\alpha \geq 1$, the contour lines will now \textit{bend outwards}.
During optimization of the Power Loss, the gradient of the loss contour map can be calculated as $-\nabla [-(G_1^\alpha + G_2^\alpha)] = (\frac{\partial (G_1^\alpha + G_2^\alpha) }{\partial G_1} , \frac{\partial (G_1^\alpha + G_2^\alpha) }{\partial G_2} )^T = (\alpha G_1^{\alpha-1},\alpha G_2^{\alpha-1})^T$, which will prioritize the optimization of higher gating values more, giving them a larger gradient magnitude and pushing them to reach $\tau$.

We visualize the case $\alpha=2$ and $\alpha=3$ in Fig. \ref{Figure.VisualizationOfGradient}.C, 
where the gradient arrows are now pointing more towards the the right or top edge, whichever is closer. This allows the depicted points in the figure to optimize a gating value to reach $\tau$, while within budget $B$.
We note that a larger $\alpha$ deforms the contour lines more, which makes the gradients directed more horizontally or vertically.
In the extreme case,  e.g., $\alpha=\infty$, the gradient direction is either vertical or horizontal. 
We conduct ablation experiments to find an optimal $\alpha$ for our method.

\subsection{Rectified Direction of Gradient}
\label{sec:RectifiedDirection}

In the previous section, our Power Loss balances the optimization between all the \textit{deactivated gates}, prioritizing the loss magnitudes of some deactivated gates over others.
In this section, instead of focusing on deactivated gates only, we look at how we can optimize the deactivated gates while \textit{not interfering with the activated gates}.
This aims to mitigate the difficulties in jointly optimizing many gates at once.
For example, there are 32 gates in SkipNet \cite{Wang_2018_ECCV} that dynamically determine which of the 32 layers to skip or not skip, and it can be difficult to jointly optimize and keep many gates activated simultaneously.
In this case, the joint optimization using the Complexity Loss might lead to a scenario where, after a backpropagation step, we manage to activate some gates but cause some others to become deactivated.
We attribute such optimization difficulties to the presence of conflicting gradients between the gates, which makes it hard to activate new gates without interfering with already-activated gates, resulting in a weaker attack.
Below, we first explain the problem of gradient conflicts in more detail, before proposing our method to tackle it.

\begin{figure*}[tp]
    \center
    \includegraphics[width=\textwidth]{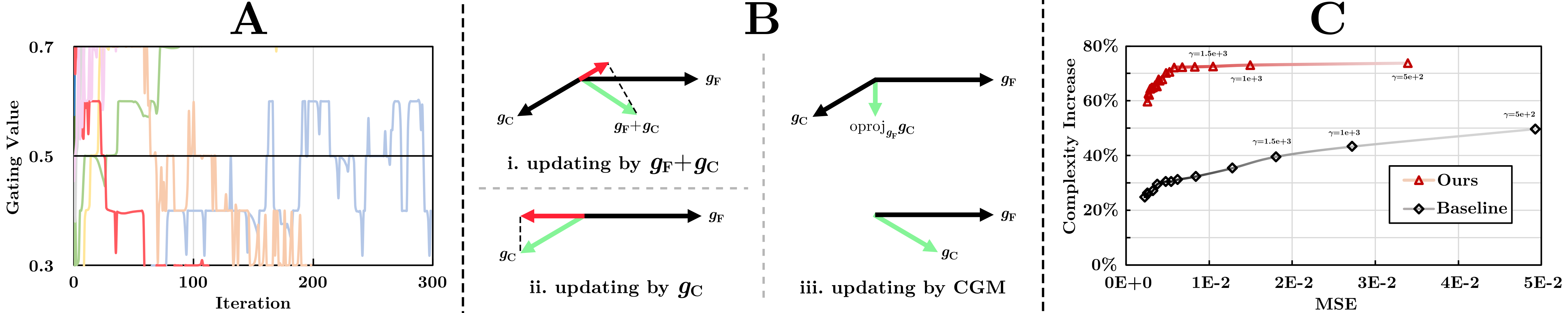}     
    \caption{ 
    A: Visualization of various gating values when optimizing using Eq.~\ref{Eq.StationarityFinal} to attack SkipNet.
    B: Illustration of gradient conflicts. (Top left) Optimizing according to Eq.~\ref{Eq.LagrangianRelaxation}. (Bottom Left) Optimizing according to Eq.~\ref{Eq.StationarityFinal}. (Right) Applying our CGM according to Eq.~\ref{Eq.RectifiedGradient}.
    C: Visualization of the increased number of activated gates at lower MSE values using our GradMDM.     
    }
    \label{Figure.Norm}
    \vspace{-.5cm}
\end{figure*}

\noindent \textbf{Gradient Conflicts}
can occur between the gradients meant to attack different gates. 
Let us represent a gradient that attacks the deactivated gates (gates with value below the threshold $\tau$) as $\boldsymbol{g}_\text{C}$, and a gradient that attacks the already-activated gates (gates with values above the threshold $\tau$) as $\boldsymbol{g}_\text{F}$.
We call $\boldsymbol{g}_\text{C}$ the Complexity Gradient, and $\boldsymbol{g}_\text{F}$ the Finished Gradient.
In the Complexity Loss defined in Eq.~\ref{Eq.StationarityFinal}, only gradients $\boldsymbol{g}_\text{C}$ from gates which are deactivated are applied for parameter updating.
However, as illustrated in Fig.~\ref{Figure.Norm}.B, $\boldsymbol{g}_\text{C}$ might be in a largely conflicting direction from $\boldsymbol{g}_\text{F}$, which might result in a higher tendency for some of the activated gates to become deactivated again.
This kind of disagreement will drive the gates to switch back and forth between being activated and deactivated.
We show an example by focusing on the gating values of various gates of SkipNet in Fig.~\ref{Figure.Norm}.A.  
Gates that have already been activated, can easily become deactivated, making it difficult to jointly activate them all at once.

Furthermore, this behaviour in the gradients cannot be simply solved by updating using both the Complexity Gradient $\boldsymbol{g}_\text{C}$ and the Finished Gradient $\boldsymbol{g}_\text{F}$.
If we jointly use all gradients as in Eq.~\ref{Eq.LagrangianRelaxation}, we can see in Fig.~\ref{Figure.Norm}.B that the gradient conflict might result in the joint gradient being in a very different direction from $\boldsymbol{g}_\text{C}$, which might obstruct the activation of other currently deactivated gates.
In addition, encouraging higher activation values for those already-activated gates by including $\boldsymbol{g}_\text{F}$ does not further contribute to the complexity increase, yet also reduces the focus on perturbation minimization as discussed in Sec.~\ref{The Overall Objective for Computational Complexity Attack}.

\noindent\textbf{Complexity Gradient Masking}.
As discussed, due to the conflicts in gradients, it is difficult to jointly activate many gates at once.
One lingering question remains: should we update using gradients $\boldsymbol{g}_\text{F}$ during updating, or not?
It seems like we are made to choose between 1) constantly trying to activate more deactivated gates which might lead to deactivation of already-activated gates and 2) prioritizing to keep the activated gates while focusing less on attacking deactivated gates.
However, we can tackle the issue with the following insight: To prevent de-activation of activated gates, we need not update using $\boldsymbol{g}_\text{F}$ to further attack already-activated gates. Instead, we only need to prevent the Complexity Gradient $\boldsymbol{g}_\text{C}$ from \textit{negatively affecting the already-activated gates}.
To this end, we propose a Complexity Gradient Masking (CGM) technique that resolves this gradient conflict issue,
such that more gates can be activated without interfering with the already-activated gates.
Firstly, we compute the Finished Loss $\mathcal{L}_\text{F}$ for those activated gates. 
The formula of the Finished Loss is similar to the Complexity Loss $\mathcal{L}_\text{C}$ in Eq.~\ref{Eq.StationarityFinal}, and is as follows: 
\vspace{-0.1cm}
\begin{align}
    \mathcal{L}_\text{F}(\boldsymbol{x}_{0},\boldsymbol{\delta})=- \sum_{i=1}^N  \max(G_i(\boldsymbol{x}_0,\boldsymbol{\delta}),\tau),
    \label{eq:Finished_Loss}
\end{align}
where as compared to Eq.~\ref{Eq.StationarityFinal}, $\min$ is changed to $\max$ to change the focus of the loss to the activated gates.
Using the Finished Loss $\mathcal{L}_\text{F}(\boldsymbol{x}_{0},\boldsymbol{\delta})$, we can compute the Finished Gradient $\boldsymbol{g}_\text{F}\in \mathbb{R}^{3\times H\times W}$ as $\boldsymbol{g}_\text{F}=\nabla_{\boldsymbol{\delta}} \mathcal{L}_\text{F}(\boldsymbol{x}_{0},\boldsymbol{\delta})$,
where $\boldsymbol{g}_\text{F}$ is the direction to optimize the perturbation $\boldsymbol{\delta}$ to further increase the gate value for activated gates.
Next, we can use either the Complexity Loss $\mathcal{L}_\text{C}$ (in Eq.~\ref{Eq.StationarityFinal}) or the Power Loss $\mathcal{L}_\text{P}$ (in Eq.~\ref{eq.power_loss}) to calculate the Complexity Gradient $\boldsymbol{g}_\text{C}\in \mathbb{R}^{3\times H\times W}$, which is the gradient direction to increase gating values for currently deactivated gates. 
If we use the Complexity Loss $\mathcal{L}_\text{C}$, $\boldsymbol{g}_\text{C}$ is computed as:
\begin{align}
        \boldsymbol{g}_\text{C}=\nabla_{\boldsymbol{\delta}} \mathcal{L}_\text{C}(\boldsymbol{x}_{0},\boldsymbol{\delta})=\nabla_{\boldsymbol{\delta}}  \sum_{i=1}^N - \min(G_i(\boldsymbol{x}_0,\boldsymbol{\delta}),\tau) .
        \label{eq:complexity_gradient}
\end{align}
Thirdly, we project the Complexity Gradient onto the Finished Gradient using:
\begin{align}
    \text{proj}_{\boldsymbol{g}_\text{F}} \boldsymbol{g}_\text{C}=\frac{\left\langle\boldsymbol{g}_\text{C}, \boldsymbol{g}_\text{F}\right\rangle}
    {{\left\lVert\boldsymbol{g}_\text{F}\right\rVert_2}}
    \frac{\boldsymbol{g}_\text{F}}{\left\lVert\boldsymbol{g}_\text{F}\right\rVert_2},
    \label{eq:gradient_projection}
\end{align}
and accordingly calculate the rejection of the Complexity Gradient from the Finished Gradient as:
\begin{align}
    \text{oproj}_{\boldsymbol{g}_\text{F}} \boldsymbol{g}_\text{C}=\boldsymbol{g}_\text{C}-\text{proj}_{\boldsymbol{g}_\text{F}} \boldsymbol{g}_\text{C},
\end{align}
whose direction is orthogonal to the Finished Gradient. 
Hence updating with it leads to less negative impact on the gates that have already been activated.
Finally, we rectify the direction of Complexity Gradient as:
\begin{align}
\label{Eq.RectifiedGradient}
    \boldsymbol{g}'_\text{C}
    =\left\{
        \begin{array}{lll}
             \boldsymbol{g}_\text{C},&&\left\langle\text{proj}_{\boldsymbol{g}_\text{F}}\boldsymbol{g}_\text{C},\boldsymbol{g}_\text{F}\right\rangle\ge 0,\\
            \text{oproj}_{\boldsymbol{g}_\text{F}} \boldsymbol{g}_\text{C},&&\left\langle\text{proj}_{\boldsymbol{g}_\text{F}}\boldsymbol{g}_\text{C},\boldsymbol{g}_\text{F}\right\rangle<0
        \end{array}
    \right.,
\end{align}
which indicates that when the projection is opposite to the direction of Finished Gradient, it will be removed and only the orthogonal component of the Complexity Gradient will be updated.

Fig.~\ref{Figure.Norm}.B visualizes the change of the angle $\theta$ between the Finished Gradient and the Complexity Gradient when our CGM method is used during training.
When the angle is obtuse, the Complexity Gradient conflicts with the Finished Gradient, i.e., updating along the original Complexity Gradient has a high risk of deactivating the activated gates. Thus, our method rectifies the direction towards a direction that is orthogonal to the Finished Gradient, which mitigates the negative impact on the activated gates.
When the angle is acute, the Complexity Gradient and Finished Gradient do not conflict, and the vanilla Complexity Gradient is used.
Overall, this allows CGM to activate more gates and effectively avoid deactivating already-activated gates.

\subsection{Training details}
\label{sec:training_details}

The techniques as described in \ref{sec:RectifiedMag} and \ref{sec:RectifiedDirection} can be applied independently to optimize perturbations. 
When only Power Loss is applied, the perturbations are optimized using $\gamma \mathcal{L}_\text{MSE} + \mathcal{L}_\text{P}$.
When only CGM is applied, Eqs. \ref{eq:Finished_Loss}-
\ref{Eq.RectifiedGradient} are computed, and $\boldsymbol{g}'_\text{C} + \nabla_{\boldsymbol{\delta}} \gamma \mathcal{L}_\text{MSE}$ is used for updating.
Using the complete GradMDM method, when both techniques are used together, Eq.~\ref{eq:Finished_Loss} is first computed, then Power Loss $\mathcal{L}_\text{P}$ is used to compute $\boldsymbol{g}_\text{C}$ as $ \boldsymbol{g}_\text{C}=\nabla_{\boldsymbol{\delta}} \mathcal{L}_\text{P}(\boldsymbol{x}_{0},\boldsymbol{\delta})$, before computing Eqs. \ref{eq:gradient_projection}-\ref{Eq.RectifiedGradient} as usual to obtain $\boldsymbol{g}'_\text{C}$. 
Lastly, $\boldsymbol{g}'_\text{C} + \nabla_{\boldsymbol{\delta}} \gamma \mathcal{L}_\text{MSE}$ is used to update the perturbations.

Additionally, in order to balance the importance of different gates according to their corresponding computational complexities, we introduce in $\mathcal{L}_{\text{C}}$ (Eq.~\ref{Eq.StationarityFinal}), $\mathcal{L}_{\text{P}}$ (Eq.~\ref{eq.power_loss}) and $\mathcal{L}_{\text{F}}$ (Eq.~\ref{eq:Finished_Loss}) weights $\{\lambda_i\}_{i=1}^N$ 
that reweigh the losses of corresponding gates $\{G_i\}_{i=1}^N$  according to their complexities. 
We define $\lambda_i$ as: $\lambda_i = \frac{C_i}{\sum_{j=1}^N C_j}$, where $C_i$ denotes the computational complexity (in GFLOPs) of the $i$-th gate's optional operations, i.e., $\lambda_i$ denotes the complexity proportion of $i$-th gate's operations to all the optional operations in the network.

\noindent\textbf{Implementation details.} 
We set our $\alpha=4$, $\gamma=100$ and
run 100 training iterations for our method.
The first 20\% of training iterations are warm-up iterations (e.g., using Eq.~\ref{Eq.StationarityFinal}) to get our performance closer to the Pareto Frontier, and the rest are GradMDM iterations.
All the original tested dynamic networks set the activation threshold at $\tau=0.5$, which we follow.
Our code is implemented using the Pytorch framework.
We use the official publicly available code for all the dynamic models.
As pre-trained models are publicly available for SkipNet and SACT on ImageNet and CIFAR-10 datasets, we directly run our attacking experiments on them.
For the other dynamic models and datasets, pre-trained models are not available, so we train them ourselves for each dataset by referring to the training details and hyperparameters in their respective works.

\section{Experiments}
We validate our approach by using it to attack popular dynamic depth networks (SkipNet \cite{Wang_2018_ECCV} and SACT \cite{figurnov2017spatially}) and a dynamic width network (ManiDP \cite{tang2021manifold}) on ImageNet \cite{deng2009imagenet} and CIFAR-10 \cite{Krizhevsky09cifar10}.

\noindent\textbf{Dataset Settings.}
For evaluation, we follow the datasets and settings of previous works \cite{Haque_2020_CVPR}.
We evaluate on ImageNet \cite{deng2009imagenet} and CIFAR-10 \cite{Krizhevsky09cifar10}, where images from ImageNet and CIFAR-10 are converted into 3$\times$224$\times$224 and 3$\times$32$\times$32, respectively.

\noindent\textbf{Metrics.}
We follow the metrics used in previous works \cite{Haque_2020_CVPR} to evaluate the effectiveness of attacks. 
We measure the amount of computation savings of the dynamic network that are invalidated by our attacks, and report the \textbf{Average Recovery Percentage (ARP)} of the reduced FLOPs during inference. 
To measure the quality of the perturbed image, we adopt the \textbf{Peak Signal-to-Noise Ratio (PSNR)} metric,
which approximates the human perceptual quality and is commonly employed to evaluate image quality \cite{hore2010image}. 
The \text{PSNR} can be defined as $\text{PSNR} = 10 \cdot \log_{10}{(\frac{\text{MAX}_\text{I}^2}{\text{MSE}})} $,
where $\boldsymbol{\text{MAX}_\text{I}}$ is the maximum possible pixel value of the image. For an original $\text{m}\times \text{n}$ image \text{I} and the perturbed image \text{K}, \textbf{Mean Squared Error (MSE)} can be calculated by $ \text{MSE}=\frac{1}{3HW}\big\lVert \boldsymbol{x}'_0-\boldsymbol{x}_0 \big\lVert^2_2$.

\subsection{Attack on Dynamic Depth Network}

We compare our approach with state-of-the-art methods ILFO \cite{Haque_2020_CVPR} and DeepSloth \cite{Hong2021DeepSloth}. 
For a fair comparison, we follow the experimental settings in \cite{Haque_2020_CVPR}.

\noindent\textbf{Model Settings.} We attack SkipNet \cite{Wang_2018_ECCV}, a conditional skipping network, with two different settings: \textbf{SkipNet} and \textbf{SkipNet+HRL}, where +HRL indicates that hybrid reinforcement learning is used. 
SkipNet+HRL achieves better efficiency compared to SkipNet and is more challenging to attack. 
We also attack \textbf{SACT} \cite{figurnov2017spatially}, an early termination network.
Since ILFO only reported results on SkipNet and SACT, we further re-implement their code on SkipNet+HRL to compare it with our GradMDM.
Note that ILFO uses a similar loss to the baseline described in Eq.~\ref{Eq.StationarityFinal}, except that their formulation instead uses different pre-set thresholds for different gates.
DeepSloth \cite{Hong2021DeepSloth} is originally designed for attacking early termination-based networks only, thus we report its results on SACT only.

\noindent\textbf{Results.} We present the results on ImageNet in Table \ref{table:dynamicDepth} and on CIFAR-10 in Table \ref{table:dynamicDepth_cifar}. 
On both CIFAR-10 and ImageNet, our GradMDM outperforms the ILFO baseline on all tested models across all metrics. 
This implies that our GradMDM attacks consistently achieve significantly higher increase in computational complexity, while using a less imperceptible perturbation (i.e., smaller MSE and larger PSNR).
Specifically, for SkipNet and SACT, GradMDM can almost achieve 100\% average recovery, which means that the dynamic networks have almost the same computational complexity as their static counterparts, which shows its effectiveness.

\begin{table}[ht]
  \vspace{-0.4cm}
  \tiny
  \centering
  \caption{Performance comparison of dynamic depth networks on ImageNet. 
  }
  \vspace{-0.2cm}
  \label{table:dynamicDepth}
  \setlength{\tabcolsep}{0.9mm}
  \setlength{\aboverulesep}{0pt}
  \setlength{\belowrulesep}{0pt}  
  \begin{tabular}{l| ccc |ccc|ccc}
    \hline
     \multirow{2}{*}{Model} & \multicolumn{3}{|c}{SkipNet} & \multicolumn{3}{|c}{SACT} & \multicolumn{3}{|c}{SkipNet+HRL} \\
     &ARP(\%)$\uparrow$ &MSE$\downarrow$&PSNR$\uparrow$&ARP(\%)$\uparrow$ &MSE$\downarrow$&PSNR$\uparrow$&ARP(\%)$\uparrow$ &MSE$\downarrow$&PSNR$\uparrow$\\
    \hline
    ILFO\cite{Haque_2020_CVPR}(baseline)  & 81.4 & 0.26 & 54.0& 91.1 & 0.25 & 54.1 & 49.6 & 0.25 & 54.3 \\
    DeepSloth\cite{Hong2021DeepSloth}  & - & - & - & 90.7  & 0.80  & 49.1  & - & - & - \\    
    GradMDM(ours) & \textbf{99.3} & \textbf{0.08} & \textbf{59.1} & \textbf{99.0} & \textbf{0.10} & \textbf{57.9} & \textbf{73.4} & \textbf{0.11} & \textbf{57.5} \\
    \bottomrule
  \end{tabular}
\end{table}

\begin{table}[ht]
  \vspace{-0.5cm}
  \tiny
  \centering

  \caption{Performance comparison of dynamic depth networks on CIFAR-10. 
  }
  \vspace{-0.2cm}
  \label{table:dynamicDepth_cifar}
  \setlength{\tabcolsep}{0.9mm}
  \setlength{\aboverulesep}{0pt}
  \setlength{\belowrulesep}{0pt}  
  \begin{tabular}{l |ccc |ccc| ccc}
    \hline
     \multirow{2}{*}{Model} & \multicolumn{3}{c}{SkipNet} & \multicolumn{3}{|c}{SACT} & \multicolumn{3}{|c}{SkipNet-HRL} \\
     &ARP(\%)$\uparrow$&MSE$\downarrow$&PSNR$\uparrow$&ARP(\%)$\uparrow$&MSE$\downarrow$&PSNR$\uparrow$&ARP(\%)$\uparrow$&MSE$\downarrow$&PSNR$\uparrow$\\
    \hline
    ILFO\cite{Haque_2020_CVPR}(baseline) & 84.3 & 0.24 & 54.3& 72.5 & 0.25 & 54.1 &5.2&0.24& 54.4\\
    DeepSloth\cite{Hong2021DeepSloth}  & - & - & - & 92.5 & 0.92  & 48.5 &-&-&-\\        
    GradMDM(ours) & \textbf{99.1} & \textbf{0.11} & \textbf{57.6} & \textbf{99.0} & \textbf{0.10} & \textbf{58.0}  & \textbf{30.0} & \textbf{0.11} & \textbf{57.4}\\
    \bottomrule
  \end{tabular}
 \vspace{-0.1cm} 
\end{table}

\subsection{Attack on Dynamic Width Network}

We next evaluate our attack on the dynamic width network ManiDP \cite{tang2021manifold}. 
We evaluate our approach on ImageNet and CIFAR-10, with results shown in Table~\ref{table:dynamicWidth}. 
Our GradMDM outperforms ILFO on both datasets on all metrics, showing its efficacy on dynamic width structures as well.

\begin{table}[ht]
  \vspace{-0.3cm}
  \centering

  \caption{Performance comparison of ManiDP on CIFAR-10 and ImageNet.  }
  \label{table:dynamicWidth}
  \vspace{-0.15cm}
  \scalebox{0.95}{
    \hspace{-0.4cm}
  \setlength{\tabcolsep}{1mm}
  \setlength{\aboverulesep}{0pt}
  \setlength{\belowrulesep}{0pt}  
  \begin{tabular}{l |ccc |ccc}
    \hline
     \multirow{2}{*}{Dataset} & \multicolumn{3}{c}{CIFAR-10} & \multicolumn{3}{|c}{ImageNet} \\
     &ARP(\%)$\uparrow$&MSE$\downarrow$&PSNR$\uparrow$&ARP(\%)$\uparrow$&MSE$\downarrow$&PSNR$\uparrow$\\     
    \hline
    ILFO\cite{Haque_2020_CVPR} (baseline) & 80.3 & 0.26 & 54.0 & 85.6 & 0.26& 54.0\\
    GradMDM (ours) & \textbf{99.0} & \textbf{0.11} & \textbf{57.5 } & \textbf{99.3} & \textbf{0.11} & \textbf{57.7} \\
    \bottomrule
  \end{tabular}
  }
\vspace{-0.3cm}
\end{table}

\subsection{Ablation Studies}

We conduct ablation studies on ImageNet using SkipNet+HRL to further investigate GradMDM.

\noindent \textbf{Components of GradMDM.} 
Tab.~\ref{table:ablation_methods} evaluates the gains brought by each individual component.
Each individual component of our method, CGM and the Power Loss (PL), can attain better attacking performance as compared to the baselines that use Eq.~\ref{Eq.LagrangianRelaxation} or Eq.~\ref{Eq.StationarityFinal} for optimization. Together, PL and CGM form our full method GradMDM, and perform better when used together.

\begin{table}[ht]
  \vspace{-0.2cm}
  \scriptsize
  \centering
  \caption{
  Ablation of individual components of GradMDM.  
  PL and CGM denote Power Loss and Complexity Gradient Masking, respectively.
  GradMDM uses the combination of both PL and CGM. 
  }
  \vspace{-0.15cm}
  \label{table:ablation_methods}
  \scalebox{0.95}{
  \hspace{-0.4cm}
  \setlength{\tabcolsep}{2mm}
  \begin{tabular}{lccccc}
    \toprule
    Method & Eq. \ref{Eq.LagrangianRelaxation}  & Eq. \ref{Eq.StationarityFinal} & CGM (Ours) & PL (Ours) & GradMDM (Ours) \\
    \midrule
    Recovery (\%) $\uparrow$ & 48.5 & 49.6 & 54.0 & 69.0 & 73.4 \\
    MSE $\downarrow$ & 0.25  & 0.25 & 0.11 & 0.11 & 0.11 \\    
    PSNR $\uparrow$ & 54.3  & 54.3 & 57.5 & 57.6 &  57.5 \\       
    \bottomrule
  \end{tabular}
  }
\end{table}

\noindent \textbf{Efficiency.} 
In Tab.~\ref{table:efficiency} we evaluate the time cost of performing GradMDM. Experiments are conducted on a Nvidia RTX 3090 GPU.
On all models, our GradMDM conducts attacks faster than ILFO.

\begin{table}[ht]
  \vspace{-0.2cm}
  \tiny
  \centering
  \caption{ Efficiency comparison of different methods on ImageNet.
  }
  \vspace{-0.15cm}
  \label{table:efficiency}
  \setlength{\tabcolsep}{0.8mm}
  \setlength{\aboverulesep}{0pt}
  \setlength{\belowrulesep}{0pt}  
  \begin{tabular}{l |ccc |ccc |ccc}
    \hline
     \multirow{2}{*}{Model} & \multicolumn{3}{|c}{SkipNet} & \multicolumn{3}{|c}{SACT} & \multicolumn{3}{|c}{SkipNet+HRL} \\
     &ARP(\%)&Iteration&Time (s) &ARP(\%)&Iteration&Time (s)&ARP(\%)&Iteration&Time (s)\\
    \hline
    ILFO\cite{Haque_2020_CVPR}(baseline)  & 81.4 & 300 & 2.1 & 91.1 & 300 & 1.8  & 49.6 & 300 & 2.0 \\
    DeepSloth\cite{Hong2021DeepSloth}  & - & - & - &  90.7 & 550 & 3.8 & - & - & - \\        
    GradMDM(ours)& \textbf{99.3} & 100 & \textbf{1.2} & \textbf{99.0} & 100 & \textbf{1.1} & \textbf{73.4} & 100& \textbf{1.1} \\    
    \bottomrule
  \end{tabular}
    
\end{table}

\noindent \textbf{Weight of Imperceptibility Loss $\gamma$.}
Tab.~\ref{table:ablation_lambda} evaluates the attacking performance and the deviation of modified input images under different settings of $\gamma$.
As expected, when $\gamma$ decreases, the efficacy of our attack improves, but the magnitude of the perturbation also increases.
When $\gamma$ is set to a larger value of $1e+4$, a very high PSNR (73.9) and low MSE (3e-3) is obtained while still simultaneously improving upon the baseline attacking performance (57.5\% as compared to baseline of 49.6\%).
To better compare between GradMDM and ILFO, we run both methods at more $\gamma$ settings and show the results in Fig. \ref{Figure.Norm}.C, where our method consistently attains a better Complexity-MSE trade-off.

\begin{table}[ht]
  \centering
  \vspace{-0.2cm}
  \caption{Ablation of the different settings of $\gamma$. The higher the value of $\gamma$, the more weight will be given to the Imperceptibility Loss in comparison to the Complexity Loss. }
  \vspace{-0.15cm}
  \label{table:ablation_lambda}
  \scalebox{0.95}{
  \hspace{-0.4cm}
  \setlength{\tabcolsep}{2mm}
  \begin{tabular}{lcccccc}
    \toprule
    $\gamma$ & 1e+0 & 1e+1 & 1e+2  & 1e+3 & 1e+4\\
    \midrule
    ARP (\%) $\uparrow$ & 76.1 & 74.5 & 73.4 & 72.0 & 57.5 \\
    MSE $\downarrow$ & 0.25 & 0.21 & 0.11 & 0.03 & 3e-3 \\
    PSNR $\uparrow$ & 54.3 & 54.7 & 57.5 & 63.8 & 73.9  \\
    \bottomrule
  \end{tabular}
  }
\end{table}

\noindent \textbf{Classification Accuracy.}
We also investigate the change in image classification accuracy after our attack,
and report the results for several dynamic network methods on ImageNet (corresponding to Tab.~\ref{table:dynamicDepth}) in Tab.~\ref{table:dynamicDepth_accuracy}. 
We find that the perturbations by GradMDM lead to a significant drop in model accuracy. 
However, during our attack, we can also add a classification loss (\textbf{GradMDM w/ classification loss}), which we find helps to improve the classification accuracy while maintaining the attack performance.

\begin{table}[ht]
\vspace{-0.3cm}
\centering
\tiny
\caption{
Classification Accuracy on ImageNet after the attack. 
  }
\vspace{-0.15cm}
  \label{table:dynamicDepth_accuracy}
  \setlength{\tabcolsep}{1.5mm}  
  \begin{tabular}{c | cc |cc| cc }
    \hline
     \multirow{2}{*}{Setting} & \multicolumn{2}{|c}{SkipNet} & \multicolumn{2}{|c}{SACT} & \multicolumn{2}{|c}{SkipNet-HRL} \\
     &ARP(\%)&Acc(\%)&ARP(\%)&Acc(\%)&ARP(\%)&Acc(\%)\\
    \hline
     GradMDM & 99.3 & 36.2 & 99.0  & 33.8  & 73.4  & 23.1 \\    
     GradMDM w/ classification loss & 99.3 & 77.5 & 99.0  & 78.2 & 73.2  & 76.9 \\
    \hline
     Original Acc. (Upper Bound) & - & 77.5 & - & 78.2 & - & 76.9  \\
    \hline    
  \end{tabular}
\end{table}

\noindent \textbf{Power Loss hyperparameter $\alpha$.} 
In Tab. \ref{table:ablation_alpha}, we evaluate the impact of different settings of $\alpha$.
Different values of $\alpha$ all lead to improvement in performance over the baseline where $\alpha=1$. 
We find that increasing $\alpha$ past 4 does not lead to further improvements, possibly because that will focus too much on particular larger gradients. 
Another way we can set hyperparameters $\alpha$ is by setting them individually for each gate.
When we carefully choose different $\alpha$ values for different gates, our performance can indeed improve slightly, with ARP, MSE and PSNR of 74.5\%, 0.11 and 57.5.
Nevertheless, it can be difficult to tune the $\alpha$ hyperparameters in this manner for different models and datasets.
Furthermore, the performance improvement from the tuning of $\alpha$ for each gate is not very significant.
Thus, we report our results with constant $\alpha$ ($\alpha=4$) for all gates.

\begin{table}[ht]
  \centering
    \vspace{-0.2cm}
  \caption{Ablation of the different settings of $\alpha$. The higher the value of $\alpha$, the larger the deformation of the contour lines in the Power Loss.  }
  \vspace{-0.15cm}
  \label{table:ablation_alpha}
  \scalebox{0.95}{
  \hspace{-0.4cm}
  \setlength{\tabcolsep}{2mm}
  \begin{tabular}{lcccccc}
    \toprule
    $\alpha$ & 1  & 2 & 3 & 4 & 5 & 6\\
    \midrule
    ARP (\%) $\uparrow$ & 50.7 & 53.0 & 72.1 & 73.4 & 72.5 & 69.0 \\
    MSE $\downarrow$ & 0.11 & 0.11 & 0.11& 0.11 & 0.11  & 0.11  \\
    PSNR $\uparrow$ & 57.5 & 57.5 & 57.6 & 57.5 & 57.7 & 57.6  \\
    \bottomrule
  \end{tabular}
  }
\end{table}

\noindent \textbf{Performance on samples that require different ``budgets".} 
We also conduct the following experiment on SkipNet using ImageNet: we select a subset of samples that can lead to 100\% activation using the baseline (Subset A) under a controlled amount of perturbation, and a subset of samples that cannot be satisfactorily tackled by the baseline (Subset B).
We find that the ARP of GradMDM on Subset A is still 100\%, while on Subset B, GradMDM leads to significant improvements on ARP (80.9\% vs 99.0\%). This shows the efficacy of our method.

\noindent \textbf{Tanh vs projection.}
In Eq.~\ref{eq:tanh_mapping} we use the tanh formulation to map each element of our raw $(\textbf{x}_0 + \boldsymbol{\delta})$ into the feasible space $[0,1]$ for fair comparison with previous works \cite{Haque_2020_CVPR}. Alternatively, we can also directly project $(\textbf{x}_0 + \boldsymbol{\delta})$ into the feasible space $[0,1]$, in a way similar to Projected Gradient Descent (PGD). 
Using this alternative setting, our GradMDM still outperforms the baseline on ARP metric (45.0\% vs 77.5\%) with a similar MSE (0.02 vs 0.02).

\noindent \textbf{L2 vs L$\infty$.}
In Eq.~\ref{eq.OriginalObjective} we use the $L_{2}$ norm for a fair comparison against previous works \cite{Haque_2020_CVPR}.
Alternatively, following some other adversarial attack works \cite{goodfellow2014explaining}, we can also apply the $L_{\infty}$ norm.
On this setting, our experiments show that GradMDM still outperforms the baseline on ARP metric (50.0\% vs 73.2\%) with a lower MSE (0.34 vs 0.27).

\noindent \textbf{Other baselines.}
We also conduct experiments on two other baselines. 
In the \textbf{Joint (Eq.~\ref{Eq.LagrangianRelaxation})} setting, we directly use Eq.~\ref{Eq.LagrangianRelaxation} to optimize the perturbations, which is equivalent to directly updating using both the Complexity Gradient $\boldsymbol{g}_\text{C}$ and the Finished Gradient $\boldsymbol{g}_\text{F}$ in a joint manner.
The other baseline is the \textbf{$\mathcal{L}_{P} + \mathcal{L}_{F}$ (Eq.~\ref{eq.power_loss} + Eq.~\ref{eq:Finished_Loss})} setting, where we iteratively optimize the perturbation with $\mathcal{L}_{P} + \mathcal{L}_{F}$ instead of using our CGM technique.
Different from our GradMDM and ILFO \cite{Haque_2020_CVPR} baseline, these baselines directly update the perturbation using gradients from $\mathcal{L}_{F}$ during optimization, and thus tend to over-optimize the gating values of already-activated gates, and lead to sub-optimal performance.

\begin{table}[ht]
\vspace{-0.2cm}
\centering
\footnotesize
\caption{Performance comparison against other baselines.
}
\vspace{-0.15cm}
  \label{table:jointablation}
  \setlength{\tabcolsep}{1.3mm} 
  \begin{tabular}{l |ccc}
    \hline
     Method&ARP(\%)$\uparrow$&MSE$\downarrow$&PSNR$\uparrow$\\    
    \hline
    Joint (Eq. \ref{Eq.LagrangianRelaxation})  & 38.1 &0.23&54.4 \\   
    $\mathcal{L}_{P} + \mathcal{L}_{F}$ (Eq. \ref{eq.power_loss} + Eq. \ref{eq:Finished_Loss}) & 66.8 & 0.11 & 57.5 \\
    GradMDM(ours) & 73.4 & 0.11 & 57.5  \\    
    \hline
  \end{tabular}
  \vspace{-0.2cm}
\end{table}

\noindent \textbf{Other dynamic architectures.}
To further evaluate the efficacy of our proposed GradMDM attack, we also perform experiments on other dynamic architectures,
We also report results on three additional dynamic network methods: DVT \cite{wang2021not}, A-ViT \cite{yin2022avit} and Shallow-deep \cite{kaya2019shallow}.
DVT is a Transformer architecture with dynamic depth,
A-ViT is a Transformer architecture with dynamic width, and Shallow-deep is an early-exit CNN architecture with a different working mechanism from SACT.
As shown in Tab. \ref{table:additionaldynamic}, GradMDM significantly outperforms the baseline on these dynamic networks as well.

\begin{table}[ht]
\vspace{-0.2cm}
\centering
\tiny
\caption{Attack performance comparison of other dynamic networks on ImageNet.
}
\vspace{-0.15cm}
  \label{table:additionaldynamic}
  \setlength{\tabcolsep}{0.9mm} 
  \begin{tabular}{l |ccc |ccc| ccc}
    \hline
     \multirow{2}{*}{Model} & \multicolumn{3}{c}{DVT} & \multicolumn{3}{|c}{A-ViT} & \multicolumn{3}{|c}{Shallow-Deep 50\%-Confidence } \\
     &ARP(\%)$\uparrow$&MSE$\downarrow$&PSNR$\uparrow$&ARP(\%)$\uparrow$&MSE$\downarrow$&PSNR$\uparrow$&ARP(\%)$\uparrow$&MSE$\downarrow$&PSNR$\uparrow$\\
    \hline
    ILFO\cite{Haque_2020_CVPR}(baseline) & 85.0 & 0.26 & 54.1 & 92.7 & 0.11 & 57.6 & 83.7 & 0.29 & 53.6 \\    
    GradMDM(ours) & \textbf{98.3} & \textbf{
0.19} & \textbf{55.3} & \textbf{95.0} & \textbf{0.10} & \textbf{58.0} & \textbf{98.0} & \textbf{0.27}  & \textbf{53.8}  \\        
    \hline
  \end{tabular}
\end{table}

\noindent \textbf{Other datasets.}
We also report the performance of GradMDM on other popular image datasets: CIFAR-100 \cite{Krizhevsky09cifar10} and CUB-200-2011 \cite{wah2011caltech} in Tab.~\ref{table:dynamicDepth_cifar100} and Tab. \ref{table:dynamicDepth_cub} respectively.
On both of these datasets, we perform attacks on several dynamic networks (SkipNet, SACT and SkipNet-HRL). Our GradMDM consistently outperforms ILFO on all settings.

\begin{table}[ht]
\vspace{-0.3cm}
\centering
\tiny
\caption{Attack performance comparison on CIFAR-100. 
}
\vspace{-0.15cm}
  \label{table:dynamicDepth_cifar100}
  \setlength{\tabcolsep}{0.9mm} 
  \begin{tabular}{l |ccc |ccc| ccc}
    \hline
     \multirow{2}{*}{Model} & \multicolumn{3}{c}{SkipNet} & \multicolumn{3}{|c}{SACT} & \multicolumn{3}{|c}{SkipNet-HRL} \\
     &ARP(\%)$\uparrow$&MSE$\downarrow$&PSNR$\uparrow$&ARP(\%)$\uparrow$&MSE$\downarrow$&PSNR$\uparrow$&ARP(\%)$\uparrow$&MSE$\downarrow$&PSNR$\uparrow$\\
    \hline
    ILFO\cite{Haque_2020_CVPR}(baseline) & 59.4 & 0.24 & 54.3& 67.1 & 0.25 & 54.1 & 6.11 &0.48& 51.3\\
    GradMDM(ours) & \textbf{98.0} & \textbf{0.03} & \textbf{63.5} & \textbf{96.9} & \textbf{0.10} & \textbf{58.3}  & \textbf{24.5} & \textbf{0.14} & \textbf{56.7}\\
    \hline
  \end{tabular}
\end{table}

\begin{table}[ht]
\vspace{-0.55cm}
\centering
\tiny
\caption{Attack performance comparison on CUB-200-2011. }
\vspace{-0.15cm}
  \label{table:dynamicDepth_cub}
  \setlength{\tabcolsep}{0.9mm} 
  \begin{tabular}{l |ccc |ccc| ccc}
    \hline
     \multirow{2}{*}{Model} & \multicolumn{3}{c}{SkipNet} & \multicolumn{3}{|c}{SACT} & \multicolumn{3}{|c}{SkipNet-HRL} \\
     &ARP(\%)$\uparrow$&MSE$\downarrow$&PSNR$\uparrow$&ARP(\%)$\uparrow$&MSE$\downarrow$&PSNR$\uparrow$&ARP(\%)$\uparrow$&MSE$\downarrow$&PSNR$\uparrow$\\
    \hline
    ILFO\cite{Haque_2020_CVPR}(baseline) & 94.4 & 0.10 & 58.4 & 94.7 & 0.09 & 58.5& 57.2 &0.23& 54.6\\
    GradMDM(ours) & \textbf{95.9} & \textbf{0.09} & \textbf{58.7} & \textbf{96.0} & \textbf{0.08} & \textbf{59.0} & \textbf{76.6} & \textbf{0.15} & \textbf{56.5}\\
    \hline
  \end{tabular}
\vspace{-0.1cm}
\end{table}

\noindent \textbf{Transferability of Perturbations.}
We also investigate the transferability of perturbations generated by GradMDM between different models as shown in Tab. \ref{table:transferability}. 
Specifically, we evaluate two scenarios: 1) where the perturbations from SkipNet are used to attack SkipNet-HRL, and vice versa; 2) where the perturbations from SkipNet are used to attack ManiDP.
Results show that the perturbations from GradMDM are transferable to some extent, which opens up the potential of it being used as a black-box attack.

\begin{table}[ht]
\vspace{-0.2cm}
\centering
\scriptsize
\caption{
Evaluation of the transferability of GradMDM perturbations between models on ImageNet. 
}
\vspace{-0.15cm}
  \label{table:transferability}
  \setlength{\tabcolsep}{0.8mm} 
    \begin{tabular}{ccc |ccc|ccc}
    \hline
     \multicolumn{3}{c}{SkipNet$\rightarrow$SkipNet-HRL} & \multicolumn{3}{|c}{SkipNet-HRL$\rightarrow$SkipNet} & \multicolumn{3}{|c}{SkipNet$\rightarrow$ManiDP}  \\  
     ARP(\%)&MSE&PSNR&ARP(\%)&MSE&PSNR&ARP(\%)&MSE&PSNR\\
    \hline
    42.5 & 0.11  & 57.6  & 53.1 & 0.11 & 57.4 & 93.7 & 0.11 & 57.6 \\       
    \hline
  \end{tabular}
  \vspace{-0.3cm}
\end{table}

\section{Discussion of Defense Measures}

In this section, we discuss some potential defensive measures against our GradMDM attack.
A possible countermeasure is to perform adversarial training \cite{madry2017towards} to improve adversarial robustness, which is an approach that has been empirically shown to be effective against many accuracy-based attacks on static networks.
Furthermore, some other simple \textit{out-of-the-box defense} approaches such as Spatial Smoothing \cite{xu2017feature} and JPEG Compression \cite{das2017keeping} can also be employed to defend against our proposed attack by pre-processing each input image.
Here, we evaluate GradMDM against several defense measures and report the results in Tab.~\ref{table:defense}. 
We find that these defense measures are beneficial to improve robustness.

\begin{table}[ht]
\centering
\footnotesize
\caption{GradMDM attack performance comparison on ImageNet under various defense measures. 
}
\vspace{-0.15cm}
  \label{table:defense}
  \setlength{\tabcolsep}{1.3mm} 
  \begin{tabular}{l |ccc }
    \hline
    Method &   ARP(\%)\\
    \hline
    Adversarial Training [a] & 39.2 \\
    Spatial Smoothing [b] & 49.0 \\
    JPEG Compression [c] & 43.8 \\    
    \hline
    No defense measures & 73.4  \\    
    \hline
  \end{tabular}
\vspace{-0.5cm}
\end{table}

\section{Applications and Significance}
The aim of energy-oriented attacks on
dynamic networks is to increase their computation complexity, which leads to increased latency in response. 
Thus, in practical scenarios, attacks against dynamic networks can be analogous to denial-of-service attacks \cite{palmieri2015energy}.
This can be dangerous in some safety-critical applications such as self-driving vehicles, where a timely response is important \cite{hu2019dynamic}.
Such attacks are also disruptive in some cloud-based Internet-of-Things (IoT) applications \cite{Hong2021DeepSloth}, where an optional model partition is deployed on the cloud,
and network latency can be increased significantly by forcefully activating the model's computations on the cloud, leading to excessive transmissions between cloud servers and the IoT device.
Lastly, the increased energy consumption from the extra computations can cause devices to run out of battery faster, which 
can be exploited by malicious parties to disable these devices (e.g., security robots, mobile devices) \cite{martin2004denial}.

\section{Conclusion}
In this work, we introduce GradMDM, a novel energy-oriented attack that adjusts the magnitude (using a Power Loss) and direction (using CGM) of gradients to effectively find a small perturbation that activates more computation units of dynamic networks during inference.
When tested on several datasets and dynamic networks, GradMDM consistently outperforms existing works on all metrics.

\noindent
\footnotesize
\textbf{Acknowledgments.}
This work is supported by MOE AcRF Tier 2 (Proposal ID: T2EP20222-0035), National Research Foundation Singapore under its AI Singapore Programme (AISG-100E-2020-065), and SUTD SKI Project (SKI 2021\_02\_06).
This work is also supported by TAILOR, a project funded by EU Horizon 2020 research and innovation programme under GA No 952215.

\bibliographystyle{unsrt}

{\tiny
\bibliography{egbib}}

\end{document}